# Prognostic Significance of Tumor-Infiltrating Lymphocytes Determined Using Deep Learning on Colorectal Cancer Pathology Images

Anran Liu[1&], Xingyu Li[1&], Hongyi Wu[1], Bangwei Guo[2], Jitendra Jonnagaddala[3*], Hong Zhang[1*], Xu Steven Xu[4*]

[1]Department of Statistics and Finance, School of Management, University of Science and Technology of China;
[2]School of Data Science, University of Science and Technology of China;
[3]School of Population Health, UNSW Sydney, Kensington, NSW, Australia;
[4] Data Science/Translational Research, Genmab Inc., Princeton, New Jersey, USA

* Corresponding author.

& These authors contributed equally to this work.

## ABSTRACT

**Purpose**: Tumor-infiltrating lymphocytes (TILs) have a significant prognostic value in cancers. However, very few automated, deep learning-based TIL scoring algorithms have been developed for colorectal cancer (CRC).

**Methods** We developed an automated, multiscale LinkNet workflow for quantifying TILs at the cellular level in CRC tumors using H&E-stained images. The predictive performance of the automatic TIL scores ($TIL_s^{Link}$) for disease progression and overall survival was evaluated using two international datasets, including 554 CRC patients from The Cancer Genome Atlas (TCGA) and 1130 CRC patients from Molecular and Cellular Oncology (MCO).

**Results**: The LinkNet model provided outstanding precision (0.9508), recall (0.9185), and overall F1 score (0.9347). Clear dose-response relationships were observed between $TIL_s^{Link}$ and the risk of disease progression or death in both TCGA and MCO cohorts. Both univariate and multivariate Cox regression analyses for the TCGA data demonstrated that patients with high TIL abundance had a significant (approximately 75%) reduction in risk for disease progression. In both the MCO and TCGA cohorts, the TIL-high group was significantly associated with improved overall survival in univariate analysis (30% and 54% reduction in risk, respectively). However, potential confounding factors were present in the MCO dataset. The favorable effects of high TIL levels were consistently observed in different subgroups (classified according to known risk factors).

**Conclusion:** The proposed deep-learning workflow for automatic TIL quantification based on LinkNet can be a useful tool for CRC. $TIL_s^{Link}$ is likely an independent risk factor for disease progression and carries predictive information of disease progression beyond the current clinical risk factors and biomarkers. The prognostic significance of $TIL_s^{Link}$ for overall survival is also evident.

# INTRODUCTION

Tumor-infiltrating lymphocyte (TIL) level is an important biomarker for risk stratification and treatment decisions in different cancers. [1-10]. The prognostic value of TILs in the tumor microenvironment (TME) of nonmetastatic colorectal cancer (CRC) has been demonstrated in multiple studies [11-14]. The Immunoscore derived from TIL densities and spatial characterization based on IHC staining is highly predictive of CRC outcomes [13]. In addition, high TIL densities are associated with improved survival outcomes [11-14]. Furthermore, TILs may be predictive of responses to both chemotherapy and immune-checkpoint inhibitor treatment for certain cancers. Therefore, assessments of TILs are increasingly being used in clinical studies and translational research, particularly with the important role of immunotherapy in cancer care.

TIL scoring is mainly performed by pathologists, which can be prone to human errors. With recent advances in digital pathology and artificial intelligence, automated TIL quantification methods based on deep-learning algorithms have emerged. Extensive research has been conducted in the field of breast cancer, leading to multiple automated, deep-learning models for TIL quantification in breast cancer [15] based on H&E images. To date, for CRC, very few deep-learning-based TIL scoring algorithms have been reported, and only one automated model has been developed [16], in which higher TIL density was significantly associated with longer progression-free survival (PFS) in CRC.

However, the published model was based on a relatively small dataset (Yonsei: N = 180) and a subset of TCGA patients (N = 268), and the relationship of TIL densities was studied only with PFS. Therefore, the generalizability of deep learning to larger clinical studies and other clinical outcomes (e.g., overall survival) remains a question. In addition, automatic quantification of TILs in the published model was based on a tile-level lymphocyte detector (112*112-$\mu m^2$). That is, the TIL infiltration percentage was calculated as the number of lymphocyte-rich tiles (predicted as positive for lymphocytes) divided by the total number of tiles classified as tumor or tumor invasive margins of a whole-slide image (WSI). Notably, the tile-level prediction of TILs cannot provide cellular level information (i.e., true TIL counts within an image tile) and therefore can potentially overestimate the TIL area of a WSI, which may consequently lead to a biased estimation of the prognostic value of TILs for clinical outcomes.

In this study, we propose an automated deep-learning network using a novel, multiscale LinkNet [17] architecture to develop a cellular level TIL detector that can provide accurate TIL quantification based on H&E-stained WSIs. Using two large international cohorts (TCGA: N = 554; MCO: N = 1130), we evaluated the utility of TILs in terms of risk stratification by retrospectively exploring the relationships between TIL density and multiple survival outcomes (i.e., PFS and OS) in patients with CRC. Subgroup and multivariate analyses were also performed according to other clinically relevant clinicopathological variables (sex, age, pT stage, pN stage, stage, venous invasion, lymphatic invasion, BRAF, KRAS, and MSI).

## MATERIALS AND METHODS

The workflow of the proposed algorithm is illustrated in Figure 1. Briefly, (1) The tissue regions of a WSI were separated from the background using the OTSU algorithm [18]. Then, the tissue regions were split into non-overlapping patches of $256 \times 256$ pixels; (2) A tissue classifier was developed to select the tumor regions; (3) A LinkNet-based lymphocyte-segmentation model was developed to detect the cell-level lymphocytes on each color-normalized image patch for tumors; and (4) TILs were quantified for the selected patches on the WSI by calculating the overall percentage of the TIL area.

**Dataset**

*Tissue classification datasets*

Kather et al. developed two pathologist-annotated datasets (NCT-CRC-HE-100K and CRC-VAL-HE-7K) consisting of CRC image tiles of nine tissue types: adipose tissue (ADI), background (BACK), debris (DEB), lymphocytes (LYM), mucus (MUC), smooth muscle (MUS), normal colon mucosa (NORM), cancer-associated stroma (STR), and colorectal adenocarcinoma epithelium (TUM) [19]. The NCT-CRC-HE-100K dataset has 100k image patches (ADI =10.5k, BACK = 10.6k, DEB = 11.5k, LYM = 11.6k, MUC = 8.9k, MUS = 13.5k, NORM = 8.7k, STR = 10.4k, and TUM = 14.3k). The CRC-VAL-HE-7K dataset has 7180 image patches (TUM = 1,223 and non-tumor = 6,957). All the image patches were $224 \times 224$ pixels at 20X magnification. We trained the classifier model with images that were resized to $256 \times 256$ pixels.

*Lymphocyte datasets*

The Lizard dataset for CRC provides full segmentation annotations for different types of nuclei such as epithelial cells, connective tissue cells, lymphocytes, plasma cells, neutrophils, and eosinophils [20]. In total, the dataset consisted of 238 histological image regions/patches (average size of $1,016 \times 917$ pixels at $20 \times$ magnification).

*MCO and TCGA datasets*

Hematoxylin and eosin (H&E)-stained whole-slide images (WSIs) were collected from both the MCO and TCGA CRC studies. The MCO dataset consisted of patients who underwent curative resection for colorectal cancer between 1994 and 2010 in New South Wales, Australia.[21] The Cancer Genome Atlas (TCGA) public dataset includes TCGA-COAD and TCGA-READ datasets.

For patients with more than one available WSI, one slide was randomly selected for each patient. Images with annotation marks or blurs were excluded. After exclusion, 1130 WSIs from 1130 patients with MCO were included in the analysis, whereas 554 WSIs and TCGA patients were available from the TCGA database. Clinical (age, stage, and sex), molecular (*BRAF* and *KRAS* mutations), and pathological data were collected for all cases. The clinical endpoint of OS (time from entry into the study to death) was available in the MCO dataset, whereas both OS and PFS were available from TCGA data.

**Tissue classification**

Similar to the CRC tissue classification by Kather et al. [22], the NCT-CRC-HE-100K dataset was used to train a tissue-type classifier model for CRC, whereas the CRC-VAL-HE-7K dataset was used to validate the model. The overall accuracy of the tissue-type classification model was 99% for the training dataset NCT-CRC-HE-100K and 94.4% for the validation image set CRC-VAL-HE-7K [23].

**Lymphocyte segmentation**

We constructed a LinkNet-based neural network to identify lymphocytes in H&E-stained images.

The LinkNet models were trained using the Lizard CRC dataset (n = 238). Because the sample sizes of the annotated lymphocyte datasets were relatively small, similar to the previous work of Lu et al. [24], Ren et al. [25]and Redmon et al. [26], the pre-trained, lightweight ResNet18 model [27] was used as the encoder. To further improve the performance of LinkNet, we used multiscale convolutional blocks [28], a combination of $3 \times 3$ and $5 \times 5$ convolution kernels, instead of the conventional single $3 \times 3$ convolutional kernels. We adopted the first block of the ResNet18 model as the initial block and the next four blocks as the encoder. Every encoder block contained two multiscale convolutional layers, followed by a BatchNorm2d layer, ReLu layer, and shortcut connections. The four encoder blocks were followed by four decoder blocks, each of which contained a $2 \times 2$ deconvolution layer and two $3 \times 3$ convolutional layers.

Eighty percent (80%) of the images were randomly selected and used to train the LinkNet model. The remaining dataset (20%) was used to evaluate the model performance. During the training process, random flips and random cuts were used for data augmentation. Each model was trained for 100 iterations. The colors of the image patches were normalized to mitigate the impact of uneven coloring [29], thereby improving the performance of the segmentation models.

**TIL Quantification**

Our method for quantifying the TIL score is to calculate the ratio of the TIL area in valid patches. We estimated the overall TIL score for each patient by computing the percentage of TIL area as $\frac{1}{N} \sum_{1}^{N} L_i / P$. $L_i$ represents the area of lymphocytes in the ith patch (total number of pixels identified as lymphocytes) and $P$ represents the area of one patch. N represents the total number of tumor patches in the WSI.

**Survival analysis**

Survival analyses were performed to determine OS and PFS. Univariate survival analyses were performed based on log-rank tests, whereas multivariate survival analyses were performed using Cox proportional hazards modeling to estimate the hazard ratio (HR) and associated confidence intervals/p-values for the effects of TILs (as continuous variable or categorical TIL density groups) and other clinical factors. The Kaplan–Meier method was used to estimate the distribution of OS and PFS. A penalized spline (the function pspline in the R package survival) was used to characterize the dose response of the TIL effect (as a continuous variable) on different survival endpoints. The optimal cut-off was identified according to maximum rank statistics using the function surv_cutpoint in the R package survminer and was used to stratify the population into two groups (TIL high and TIL low) for categorical TIL analyses. The multivariate analysis included age, sex (male vs. female), stage (stage III-IV vs. stage I-II), lympho vascular invasion (yes vs. no), MSI (MSI-H vs. MSS), pN stage (pN2 vs. pN0-1), pT stage (pT4 vs. pT1-3), BRAF and KRAS mutations (mutated vs. WT). In the multivariate analysis, missing data were imputed with the most frequent category for each variable. All statistical analyses were performed using the R software (version 4.0.3).

# RESULTS

## Performance of LinkNet for quantification of TILs

We used the Lizard lymphocyte dataset for CRC to assess the performance of our Linknet models for lymphocyte segmentation. As expected, our LinkNet model provided outstanding precision (0.951), recall (0.919), and F1 score (0.935).

**Patient characteristics for TCGA and MCO cohorts**

A total of 1684 CRC patients were included in the analysis (Table 1): 1130 patients were from the MCO dataset and 554 patients were from the TCGA dataset [30]. The Baseline patient characteristics and demographics were similar between the MCO and TCGA studies. The median age of the MCO dataset was 69 years (range:24 – 99 years), whereas the median age of the TCGA dataset was 66 years (range: 31–90 years). Of the MCO and TCGA patients, 55% and 47%, respectively, were male. The MCO population consisted of 53% Stage I/II patients and 47% Stage III/IV patients, whereas the TCGA population had 46% and 40% Stage I/II and III/IV patents, respectively. Stage data were missing in 14% of the TCGA patients. In addition, 16% and 11% of subjects from MCO and TCGA, respectively, had MSI-H status.

After a median follow-up of 59 months, 413 events (deaths) were recorded in the MCO dataset, while 117 deaths and 149 progressions were recorded in the TCGA dataset following a median follow-up of 24.3 months. The median OS was 22.7 months (95% confidence interval (CI), 1–108 months) for TCGA patients, while the median survival was not reached for MCO patients at 59 months. The median time to progression in TCGA dataset was 20 months (95% CI, 1–101 months).

**Lymphocyte infiltration characteristics**

$TIL_S^{Link}$ was characterized by the TIL infiltrate percentage. The mean percent infiltration was 0.159% (min: 0.031%, max: 0.964%) for the TCGA CRC cohort and 0.097% (min: 0.002%, max: 0.595%) for the MCO cohort (Table 2). Overall, the TIL infiltration percentage ($TIL_S^{Link}$ scores) appeared to be higher in the TCGA population than in the MCO population. We also characterized the differences in $TIL_s^{Link}$ scores in both TCGA and MCO cohorts according to known clinical and molecular risk factors (Table 2). Significantly higher TIL infiltration (p < 0.05) was observed in patients without lymphatic invasion in both TCGA and MCO cohorts. In addition, higher TIL levels were associated with microsatellite instability in both the TCGA and MCO cohorts. This observation is in line with existing literature, where high proportions of tumor-infiltrating lymphocytes are strongly associated with microsatellite instability [31]. For MCO, a significantly higher TIL percentage was observed in pT1-3 ($P < 0.001$), pN0-1 ($P = 0.028$), TNM Stage I/II ($P = 0.035$), BRAF mutants ($P = 0.038$), and no venous invasion ($P = 0.003$).

**Prognostic Evaluation of $TIL_s^{Link}$**

***Dose-response between outcomes and $TIL_s^{Link}$***

First, the clinical outcomes (PFS and OS) were analyzed using the penalized spline model, which characterizes a non-linear dose-response association between TILs (as a continuous variable) and an outcome. Overall, as expected, the pspline model showed that the risk of disease progression or death decreased with increasing TIL levels in both the TCGA and MCO cohorts. An approximately linear dose-response relationship was observed between the risk of death and TILs in MCO. Overall, the higher TIL levels, the lower risk of death. However, for both PFS and OS in the TCGA-CRC cohort, non-linear relationships were observed. TCGA patients with TIL abundance smaller than ~0.2% demonstrated a relatively flat dose-response relationship and had a high risk of death or disease progression, whereas TCGA patients with TILs > ~ 0.2% showed a decreasing dose-response.

***Optimal TIL cutoff***

We stratified TILs into two prognostic groups, TIL-high and TIL-low, using maximum rank statistics for optimal cutoff selection. The optimal cutoff for TCGA dataset was 0.25%, while the optimal cutoff for MCO was 0.12%. The groups based on the optimal cutoffs were used to estimate PFS/OS according to the Kaplan–Meier method. The proportion of patients classified as TIL-High was 11% in TCGA and 29% in MCO.

### $TIL_s^{Link}$ Associated with Improved PFS and OS in CRC

#### _Univariate Analysis_

Univariate Cox proportional hazards models were performed for PFS in TCGA cohort and OS in TCGA and MCO cohorts, separately. Univariate analyses revealed that the $TIL_s^{Link}$ score was significantly associated with improved PFS (Figure 3) and OS (Figure 4) in patients with CRC. The TIL-high group ($TIL_s^{Link} > 0.252\%$) had a significantly longer time to progression (PFS) (hazard ratio [HR] = 0.259; 95% CI, 0.163–0.412; $P = 0.0005$) than the TIL-low group ($TIL_s^{Link} < 0.252\%$) in the TCGA cohort. Similarly, a significant trend in favor of the TIL-High group was observed for OS in both TCGA (HR = 0.464; 95% CI, 0.274–0.784; $P = 0.031$) and MCO patients (HR = 0.708; 95% CI, 0.568–0.882; $P = 0.005$) (Figure 4). That is, the TIL-High class was associated with significantly improved survival in both the TCGA and MCO datasets. Therefore, the univariate analyses revealed the significant prognostic value of our automated $TIL_s^{Link}$ not only for disease progression, but also for overall survival.

#### _Subgroup Analysis_

In subgroup analyses stratified by known clinicopathologic risk factors, favorable effects of high TIL levels on PFS in the TCGA dataset (Figure 3) were consistent across all the tested subgroups.

The hazard ratios in the subgroups ranged from 0.1 to 0.5, suggesting that the high TIL group had a 50%–90% lower risk of disease progression compared to the low TIL group. In addition, statistical significance was observed in male patients, young patients (≤ 70 years), pT1-3, pN0-1, Stage III/IV, BRAF WT, KRAS mutants, WT, and no lympho vascular invasion. For OS, in the TCGA cohort (Figure 4), a significant improvement in survival was observed in females, pN0-1, pT1-3, and BRAF WT. In the MCO cohort (Figure 5), a statistically significant improvement in OS was observed in males, young patients (age ≤ 70 years), pN0-1, MSS, Stage III/IV, KRAS, and BRAF WT. The consistent, favorable associations between high $TIL_S^{Link}$ and improved PFS and OS across different subgroups in the established risk parameters confirmed the robustness of the prognostic value of $TIL_S^{Link}$ for patients with CRC.

### _Multivariate Analysis_

In the multivariate Cox PH models (Table 3) for PFS in TCGA cohort, the prognostic value of $TIL_S^{Link}$ remained statistically significant and independent of the established clinical and molecular risk factors for CRC (i.e., pT stage, pN stage, MSI status, lympho vascular invasion, BRAF mutation, KRAS mutation, age, and sex). The TIL-high patients had a ~70% reduction in the risk of disease progression compared to the TIL-low patients (HR = 0.272; 95% CI, 0.118–0.624; $P$ = 0.002). The statistical significance of $TIL_S^{Link}$ in the multivariate setting confirms that it is a robust and independent predictor of disease progression in patients with CRC and carries predictive information of progression beyond the current clinical and biomarker risk factors.

Multivariate analysis of OS indicated a consistent effect of $TIL_S^{Link}$ on the risk of death. In TCGA cohort (Table 4), a similar decrease in the risk of death (~50%) was observed in the multivariate model after accounting for reported risk factors (HR = 0.525; 95% CI, 0.250–1.106) compared to univariate analysis (HR = 0.464; 95% CI, 0.274–0.789). The *P* value of TILs in the multivariate analysis was less significant (*P* = 0.090) than that in the univariate analysis (*P* = 0.031), which is often observed in the multivariate analysis. For the MCO cohort (Table 5), the multivariate analysis still resulted in a favorable effect for TIL-high patients, that is, a significantly lower risk of death after accounting for other risk factors. However, this effect was not statistically significant. Further examination revealed unbalanced TIL data between the pT1-3 and pT4 groups, pN0-1 vs. pN2, and UICC Stage I/II vs. III/IV (Table 2), suggesting that the change in statistical significance in the multivariate analysis is likely due to confounding. Further confirmation of the prognostic value of TILs for OS in patients with CRC using separate datasets is warranted in the future.

## DISCUSSION

Lymphocyte infiltration in the tumor microenvironment is a biomarker of anti-tumor T cell-mediated immunity. TILs are not only robustly predictive of prognosis but also inform therapeutic decision-making for immune-checkpoint inhibitors (ICI), as patients with high TIL densities may potentially benefit from ICI targeting the programmed death 1 (PD-1)/PD-L1 signaling axis, particularly in both early and advanced TNBC and HER2-positive breast cancers [32,33]. Therefore, incorporation of TILs into clinical practice is strongly considered. However, although multiple automated TIL quantification artificial intelligence model-based deep learning algorithms have been developed for breast cancers, scoring of TILs in CRC has been mainly performed by pathologists—this time-consuming process can be prone to human errors and is labor intensive. In this study, we present an automated framework based on a novel, high-resolution, multiscale LinkNet model to accurately detect TILs at the cellular level in H&E-stained WSIs from patients with CRC.

To date, most existing automated deep-learning workflows for TILs have been based on the tile-level detection of lymphocytes [33-35]. To our knowledge, only one automated deep learning-based TIL detector has been developed to automatically quantify TILs in patients with CRC. The CRC model was also based on a tile-level lymphocyte detector (112*112-$\mu m^2$). Recently, with recent advances in image segmentation, U-Net-based deep-learning models have been shown to significantly improve the accuracy of lymphocyte detection at the cellular level for breast cancers [24,36]. LinkNet was proposed to further improve the accuracy and efficiency of segmentation compared to U-Net. Similar to U-Net, LinkNet is capable of detecting TILs at the cellular level. One of the limitations of U-Net is that spatial information is lost during the decoding process sequence. LinkNet uses a link to bypass the input of the encoder and feed it into the output of the decoder to recover lost spatial information. LinkNet can provide more accurate segmentation of satellite images [17] and nuclei in H&E images than conventional CNN and U-Net. In addition, because LinkNet shares the information learned by the encoder, the decoder uses very few parameters, and LinkNet is a more efficient and lightweight algorithm that may be used for real-time segmentations.

The TIL score based on our LinkNet architecture ($TIL_s^{Link}$) is a strong prognostic factor for disease progression (PFS) in patients with CRC. There was a clear dose response between the percentage of TILs and the risk of tumor progression, and a gradual decrease in the risk with increasing TIL percentage was observed. When evaluated as a categorical variable, the TIL-high group was associated with significantly improved PFS in both univariate and multivariate analyses after accounting for known clinical and molecular risk factors for CRC, strongly indicating that $TIL_s^{Link}$ is likely to be an independent risk factor for progression and carries predictive information of disease progression beyond the current clinical and biomarker risk factors.

In both the MCO and TCGA studies, the TIL-high group according to the automated $TIL_s^{Link}$ score was significantly associated with improved overall survival based on univariate analyses. Clear dose-response relationships were also observed between the percentage of TILs and the risk of death in both studies. In TCGA CRC patients, a consistent hazard ratio was observed in both univariate and multivariate analyses after adjusting for other established risk factors, whereas in MCO patients with CRC, a reduced effect and less significant relationship were observed after adding the known risk factors in a multivariate model. Further investigation of the MCO data revealed confounding factors due to unbalanced data (TIL percentage) in different pT, pN, and UICC stage subgroups (Table 2), potentially leading to the reduced statistical significance of TILs for OS in the MCO dataset. Further studies are warranted to investigate the predictive performance of TILs for OS in CRC patients, and to understand how to integrate TILs into a multivariate prognostic model with other risk variables to optimize the survival of patients with CRC. In addition, prospective trials using baseline and/or on-treatment TILs should be considered to confirm the utility of TILs in clinical trials and real-world clinical practice.

Subgroup analyses of PFS revealed that patients with high TIL density tended to have a lower risk of disease progression across all subgroups according to known risk factors, demonstrating the robustness of the prognostic significance of our automated $TIL_s^{Link}$ for disease progression. In addition, the favorable effects on disease progression were statistically significant in males, young patients (≤ 70 years), pT1-3, pN0-1, Stage III/IV, BRAF WT, KRAS mutants/WT, and no lympho vascular invasion. Based on the OS analysis, the TIL-high group showed a statistically significant reduction in the risk of death in females, pN0-1, pT1-3, and BRAF WT for the TCGA dataset, while in males, young patients (age ≤ 70 years), pN0-1, MSS, Stage III/IV, KRAS, and BRAF WT for the MCO dataset. The prognostic value of TIL density may be dependent on primary tumor sidedness, and patients with low TIL levels and right-sided tumors tend to have a poor prognosis [14]. Our subgroup analysis suggests that the favorable effects of high TIL levels were more pronounced in young patients, pT1-3, pN0-1, Stage III/IV, and BRAF WT.

# References

1.  Adams, S.; Gray, R.J.; Demaria, S.; Goldstein, L.; Perez, E.A.; Shulman, L.N.; Martino, S.; Wang, M.; Jones, V.E.; Saphner, T.J.; et al. Prognostic value of tumor-infiltrating lymphocytes in triple-negative breast cancers from two phase III randomized adjuvant breast cancer trials: ECOG 2197 and ECOG 1199. *J Clin Oncol* **2014**, *32*, 2959-2966, doi:10.1200/JCO.2013.55.0491.

2.  Costa, R.L.B.; Gradishar, W.J. Triple-negative breast cancer: current practice and future directions. *J Oncol Pract* **2017**, *13*, 301-303, doi:10.1200/Jop.2017.023333.

3.  Foulkes, W.D.; Smith, I.E.; Reis, J.S. Triple-negative breast cancer. *New Engl J Med* **2010**, *363*, 1938-1948, doi:DOI 10.1056/NEJMra1001389.

4.  Hammerl, D.; Smid, M.; Timmermans, A.M.; Sleijfer, S.; Martens, J.W.M.; Debets, R. Breast cancer genomics and immuno-oncological markers to guide immune therapies. *Semin Cancer Biol* **2018**, *52*, 178-188, doi:10.1016/j.semcancer.2017.11.003.

5.  Hudecek, J.; Voorwerk, L.; van Seijen, M.; Nederlof, I.; de Maaker, M.; van den Berg, J.; van de Vijver, K.K.; Sikorska, K.; Adams, S.; Demaria, S.; et al. Application of a risk-management framework for integration of stromal tumor-infiltrating lymphocytes in clinical trials. *Npj Breast Cancer* **2020**, *6*, doi:ARTN 15 10.1038/s41523-020-0155-1.

6.  Plevritis, S.K.; Munoz, D.; Kurian, A.W. Association of screening and treatment with breast cancer mortality by molecular subtype in US women, 2000-2012 (vol 319, pg 154, 2018). *Jama-J Am Med Assoc* **2018**, *319*, 724-724, doi:10.1001/jama.2018.0632.

7.  Salgado, R.; Denkert, C.; Campbell, C.; Savas, P.; Nuciforo, P.; Aura, C.; de Azambuja, E.; Eidtmann, H.; Ellis, C.E.; Baselga, J.; et al. Tumor-infiltrating lymphocytes and associations with pathological complete response and event-free survival in HER2-positive early-stage breast cancer treated with lapatinib and trastuzumab: a secondary analysis of the NeoALTTO trial. *JAMA Oncol* **2015**, *1*, 448-454, doi:10.1001/jamaoncol.2015.0830.

8.  Savas, P.; Salgado, R.; Denkert, C.; Sotiriou, C.; Darcy, P.K.; Smyth, M.J.; Loi, S. Clinical relevance of host immunity in breast cancer: from TILs to the clinic. *Nat Rev Clin Oncol* **2016**, *13*, 228-241, doi:10.1038/nrclinonc.2015.215.

9.  Stanton, S.E.; Adams, S.; Disis, M.L. Variation in the incidence and magnitude of tumor-infiltrating lymphocytes in breast cancer subtypes: a systematic review. *JAMA Oncol* **2016**, *2*, 1354-1360, doi:10.1001/jamaoncol.2016.1061.

10. Stanton, S.E.; Disis, M.L. Clinical significance of tumor-infiltrating lymphocytes in breast cancer. *J Immunother Cancer* **2016**, *4*, 59, doi:10.1186/s40425-016-0165-6.

11. Eriksen, A.C.; Sorensen, F.B.; Lindebjerg, J.; Hager, H.; dePont Christensen, R.; Kjaer-Frifeldt, S.; Hansen, T.F. The prognostic value of tumor-infiltrating lymphocytes in stage II colon cancer. a nationwide population-based study. *Transl Oncol* **2018**, *11*, 979-987, doi:10.1016/j.tranon.2018.03.008.

12. Galon, J.; Costes, A.; Sanchez-Cabo, F.; Kirilovsky, A.; Mlecnik, B.; Lagorce-Pagès, C.; Tosolini, M.; Camus, M.; Berger, A.; Wind, P.; et al. Type, density, and location of immune cells within human colorectal tumors predict clinical outcome. *Science* **2006**, *313*, 1960-1964, doi:10.1126/science.1129139.

13. Pages, F.; Mlecnik, B.; Marliot, F.; Bindea, G.; Ou, F.S.; Bifulco, C.; Lugli, A.; Zlobec, I.; Rau, T.T.; Berger, M.D.; et al. International validation of the consensus Immunoscore for the classification of colon cancer: a prognostic and accuracy study. *Lancet* **2018**, *391*, 2128-2139, doi:10.1016/S0140-6736(18)30789-X.

14. Saberzadeh-Ardestani, B.; Foster, N.R.; Lee, H.E.; Shi, Q.; Alberts, S.R.; Smyrk, T.C.; Sinicrope, F.A. Association of tumor-infiltrating lymphocytes (TILs) with survival depends on primary tumor sidedness in stage III colon cancers (NCCTG N0147)

[Alliance]. *Ann Oncol* **2022**, doi:10.1016/j.annonc.2022.07.1942.

15. Loi, S.; Drubay, D.; Adams, S.; Pruneri, G.; Francis, P.A.; Lacroix-Triki, M.; Joensuu, H.; Dieci, M.V.; Badve, S.; Demaria, S.; et al. Tumor-infiltrating lymphocytes and prognosis: a pooled individual patient analysis of early-stage triple-negative breast cancers. *J Clin Oncol* **2019**, *37*, 559-569, doi:10.1200/JCO.18.01010.

16. Xu, H.; Cha, Y.J.; Clemenceau, J.R.; Choi, J.; Lee, S.H.; Kang, J.; Hwang, T.H. Spatial analysis of tumor-infiltrating lymphocytes in histological sections using deep learning techniques predicts survival in colorectal carcinoma. *J Pathol Clin Res* **2022**, *8*, 327-339, doi:10.1002/cjp2.273.

17. Chaurasia, A.; Culurciello, E. LinkNet: Exploiting encoder representations for efficient semantic segmentation. In Proceedings of the IEEE VCIP, 10-13 Dec. 2017, 2017; pp. 1-4.

18. Otsu, N. A threshold selection method from gray-level histograms. *IEEE Transactions on Systems, Man, and Cybernetics* **1979**, *9*, 62-66, doi:10.1109/TSMC.1979.4310076.

19. Li, X.; Jonnagaddala, J.; Yang, S.; Zhang, H.; Xu, X.S. A retrospective analysis using deep-learning models for prediction of survival outcome and benefit of adjuvant chemotherapy in stage II/III colorectal cancer. *J Cancer Res Clin Oncol* **2022**, *148*, 1955-1963, doi:10.1007/s00432-022-03976-5.

20. Graham, S.; Jahanifar, M.; Azam, A.; Nimir, M.; Tsang, Y.W.; Dodd, K.; Hero, E.; Sahota, H.; Tank, A.; Benes, K.; et al. Lizard: a large-scale dataset for colonic nuclear instance segmentation and classification. *Ieee Int Conf Comp V* **2021**, 684-693, doi:10.1109/Iccvw54120.2021.00082.

21. Jonnagaddala, J.; Croucher, J.L.; Jue, T.R.; Meagher, N.S.; Caruso, L.; Ward, R.; Hawkins, N.J. Integration and analysis of heterogeneous colorectal cancer data for translational research. *Stud Health Technol* **2016**, *225*, 387-391, doi:10.3233/978-1-61499-658-3-387.

22. Kather, J.N.; Krisam, J.; Charoentong, P.; Luedde, T.; Herpel, E.; Weis, C.A.; Gaiser, T.; Marx, A.; Valous, N.A.; Ferber, D.; et al. Predicting survival from colorectal cancer histology slides using deep learning: A retrospective multicenter study. *Plos Med* **2019**, *16*, e1002730, doi:ARTN e1002730
10.1371/journal.pmed.1002730.

23. Li, X.Y.; Jonnagaddala, J.; Yang, S.H.; Zhang, H.; Xu, X.S. A retrospective analysis using deep-learning models for prediction of survival outcome and benefit of adjuvant chemotherapy in stage II/III colorectal cancer. *J Cancer Res Clin* **2022**, *148*, 1955-1963, doi:10.1007/s00432-022-03976-5.

24. Lu, Z.; Xu, S.; Shao, W.; Wu, Y.; Zhang, J.; Han, Z.; Feng, Q.; Huang, K. Deep-learning-based characterization of tumor-infiltrating lymphocytes in breast cancers from histopathology images and multiomics data. *JCO Clin Cancer Inform* **2020**, *4*, 480-490, doi:10.1200/CCI.19.00126.

25. Ren, S.; He, K.; Girshick, R.; Sun, J. Faster R-CNN: towards real-time object detection with region proposal networks. *IEEE Transactions on Pattern Analysis and Machine Intelligence* **2017**, *39*, 1137-1149, doi:10.1109/TPAMI.2016.2577031.

26. Redmon, J.; Divvala, S.; Girshick, R.; Farhadi, A. You only look once: unified, real-time object detection. In Proceedings of the IEEE CVPR, 27-30 June 2016, 2016; pp. 779-788.

27. He, K.; Zhang, X.; Ren, S.; Sun, J. Deep residual learning for image recognition. In Proceedings of the IEEE CVPR, 27-30 June 2016, 2016; pp. 770-778.

28. Su, R.; Zhang, D.; Liu, J.; Cheng, C. MSU-Net: multi-scale U-Net for 2D medical image segmentation. *Front Genet* **2021**, *12*, 639930, doi:10.3389/fgene.2021.639930.

29. Anand, D.; Ramakrishnan, G.; Sethi, A. Fast GPU-enabled color normalization for digital pathology. *Int Conf Syst Signal* **2019**, 219-224.

30. Compton, C.C.; Greene, F.L. The staging of colorectal cancer: 2004 and beyond. *CA Cancer J Clin* **2004**, *54*, 295-308, doi:10.3322/canjclin.54.6.295.

31. Bilal, M.; Raza, S.E.A.; Azam, A.; Graham, S.; Ilyas, M.; Cree, I.A.; Snead, D.; Minhas, F.; Rajpoot, N.M. Development and validation of a weakly supervised deep learning framework to predict the status of molecular pathways and key mutations in colorectal cancer from routine histology images: a retrospective study. *Lancet Digit Health* **2021**, *3*, e763-e772, doi:10.1016/S2589-7500(21)00180-1.

32. Luen, S.J.; Salgado, R.; Fox, S.; Savas, P.; Eng-Wong, J.; Clark, E.; Kiermaier, A.; Swain, S.M.; Baselga, J.; Michiels, S.; et al. Tumour-infiltrating lymphocytes in advanced HER2-positive breast cancer treated with pertuzumab or placebo in addition to trastuzumab and docetaxel: a retrospective analysis of the CLEOPATRA study. *Lancet Oncol* **2017**, *18*, 52-62, doi:10.1016/S1470-2045(16)30631-3.

33. Sun, P.; He, J.; Chao, X.; Chen, K.; Xu, Y.; Huang, Q.; Yun, J.; Li, M.; Luo, R.; Kuang, J.; et al. A computational tumor-infiltrating lymphocyte assessment method comparable with visual reporting guidelines for triple-negative breast cancer. *EBioMedicine* **2021**, *70*, 103492, doi:10.1016/j.ebiom.2021.103492.

34. Fassler, D.J.; Torre-Healy, L.A.; Gupta, R.; Hamilton, A.M.; Kobayashi, S.; Van Alsten, S.C.; Zhang, Y.W.; Kurc, T.; Moffitt, R.A.; Troester, M.A.; et al. Spatial characterization of tumor-infiltrating lymphocytes and breast cancer progression. *Cancers* **2022**, *14*, 2148, doi:ARTN 2148
10.3390/cancers14092148.

35. Saltz, J.; Gupta, R.; Hou, L.; Kurc, T.; Singh, P.; Nguyen, V.; Samaras, D.; Shroyer, K.R.; Zhao, T.H.; Batiste, R.; et al. Spatial organization and molecular correlation of tumor-infiltrating lymphocytes using deep learning on pathology images. *Cell Rep* **2018**, *23*, 181-193, doi:10.1016/j.celrep.2018.03.086.

36. Thagaard, J.; Stovgaard, E.S.; Vognsen, L.G.; Hauberg, S.; Dahl, A.; Ebstrup, T.; Dore, J.; Vincentz, R.E.; Jepsen, R.K.; Roslind, A.; et al. Automated quantification of sTIL density with H&E-based digital image analysis has prognostic potential in triple-negative breast cancers. *Cancers* **2021**, *13*, 3050, doi:ARTN 3050
10.3390/cancers13123050.

## Table 1

| Table 1 Baseline patient characteristics of the TCGA and MCO CRC cohorts | | |
|---|---|---|
| | **TCGA cohort** | **MCO cohort** |
| | (n=554) | (n=1130) |
| **Age, years** | 66(31-90) | 69(24-99) |
| **Sex** | | |
| Female | 232(43%) | 509(45%) |
| Male | 263(47%) | 621(55%) |
| Missing | 59(10%) | 0 |
| **TIL** | | |
| Low | 493(89%) | 856(76%) |
| High | 61(11%) | 274(24%) |
| **pT stage** | | |
| pT1-pT3 | 439(79%) | 856(76%) |
| pT4 | 55(10%) | 274(24%) |
| Missing | 60(11%) | 0 |
| **pN stage** | | |
| N0-N1 | 402(72%) | 932(82%) |
| N2 | 92(17%) | 198(18%) |
| Missing | 60(11%) | 0 |
| **MSI** | | |
| MSI-H | 58(11%) | 157(14%) |
| MSS,MSI-L | 329(59%) | 916(81%) |
| Missing | 167(30%) | 57(5%) |
| **Stage** | | |
| Stage I-II | 256(46%) | 632(56%) |
| Stage III-IV | 220(40%) | 496(43.8%) |
| Missing | 68(14%) | 2(0.2%) |
| **BRAF** | | |
| Wild type | 290(52%) | 941(84%) |
| Mutated | 44(8%) | 132(11%) |
| Missing | 220(40%) | 57(5%) |
| **KRAS** | | |
| Wild type | 177(32%) | 724(64%) |
| Mutated | 157(28%) | 349(31%) |
| Missing | 220(40%) | 57(5%) |
| **Venous invasion** | | |
| NO | 327(59%) | 824(73%) |
| YES | 120(22%) | 281(25%) |
| Missing | 107(19%) | 25 (2%) |
| **Lymphatic invasion** | | |
| NO | 263(47%) | 702(62%) |
| YES | 201(36%) | 428(38%) |
| Missing | 90(17%) | 0 |

```
Table 2
```

| Table 2 Lymphocyte infiltration characteristics in different subgroups of the TCGA and MCO patients | | | | |
|---|---|---|---|---|
| | **TCGA cohort** | | **MCO cohort** | |
| | (n=554) | | (n=1130) | |
| | **Mean (%; SD)** | **P value*** | **Mean (SD)** | **P value*** |
| **Age** | | | | |
| **≤ 70 yrs** | **0.158(0.102)** | **0.488** | **0.097(0.049)** | **0.890** |
| **> 70 yrs** | **0.150(0.072)** | | **0.098(0.046)** | |
| Sex | | 0.363 | | 0.437 |
| Female | 0.159(0.102) | | 0.097(0.049) | |
| Male | 0.151(0.080) | | 0.097(0.046) | |
| **pN stage** | | **0.879** | | **0.028 *** |
| **N0-N1** | **0.154(0.087)** | | **0.099(0.049)** | |
| **N2** | **0.160(0.110)** | | **0.089(0.038)** | |
| pT stage | | 0.319 | | <0.001 *** |
| pT1-pT3 | 0.152(0.086) | | 0.101(0.050) | |
| pT4 | 0.156(0.093) | | 0.087(0.036) | |
| **Stage** | | **0.249** | | **0.035 *** |
| **Stage I-II** | **0.156(0.093)** | | **0.098(0.044)** | |
| **Stage III-IV** | **0.148(0.072)** | | **0.095(0.048)** | |
| KRAS | | 0.417 | | 0.619 |
| Wild type | 0.191(0.109) | | 0.098(0.049) | |
| Mutated | 0.181(0.098) | | 0.095(0.044) | |
| **BRAF** | | **0.285** | | **0.038 *** |
| **Wild type** | **0.183(0.101)** | | **0.097(0.048)** | |
| **Mutated** | **0.204(0.123)** | | **0.102(0.047)** | |
| MSI | | 0.073 . | | 0.003 * |
| MSI-H | 0.162(0.093) | | 0.104(0.050) | |
| MSS,MSI-L | 0.138(0.073) | | 0.096(0.047) | |
| **Venous invasion** | | **0.401** | | **0.003 *** |
| **NO** | **0.160(0.099)** | | **0.099(0.047)** | |
| **YES** | **0.149(0.067)** | | **0.090(0.039)** | |
| Lymphatic invasion | | 0.026 * | | 0.035 * |
| NO | 0.167(0.096) | | 0.100(0.050) | |
| YES | 0.142(0.083) | | 0.093(0.042) | |

* p values were calculated using Spearman test

# Table 3

| Table 3 Univariate and multivariate Cox regression models for progression-free survival (PFS) in TCGA patients. | | | | |
|---|---|---|---|---|
| **Univariate Cox for TCGA dataset (PFS)** | | | **Multivariate Cox for TCGA dataset (PFS)** | |
| Variable | HR(95% CI) | P value | HR(95% CI) | P value |
| **TIL Score** | | | | |
| Low | 1(ref) | - | 1(ref) | - |
| High | 0.259 (0.163, 0.412) | <0.001 *** | 0.272 (0.118, 0.624) | 0.002 ** |
| **Sex** | | | | |
| Male | 1(ref) | - | 1(ref) | - |
| Female | 0.884 (0.629,1.241) | 0.475 | 0.902 (0.644,1.264) | 0.550 |
| Age | 1.00 (0.986, 1.014) | 0.978 | 1.002 (0.987, 1.016) | 0.822 |
| **pT stage** | | | | |
| T1-T3 | 1(ref) | - | 1(ref) | - |
| T4 | 3.427 (1.794, 6.546) | <0.001 *** | 2.076 (1.332, 3.234) | 0.001 ** |
| **pN stage** | | | | |
| N0-N1 | 1(ref) | - | 1(ref) | - |
| N2 | 3.679 (2.247, 6.021) | <0.0001*** | 2.188 (1.424, 3.361) | <0.001 *** |
| **MSI** | | | | |
| MSS | 1(ref) | - | 1(ref) | - |
| MSI-H | 0.993 (0.574, 1.720) | 0.797 | 1.147 (0.631, 2.085) | 0.653 |
| **Stage** | | | | |
| Stage I-II | 1(ref) | - | 1(ref) | - |
| Stage III-IV | 2.862 (2.037, 4.021) | <0.0001*** | 1.903 (1.220, 2.970) | 0.005 ** |
| **BRAF** | | | | |
| Mutated | 1(ref) | - | 1(ref) | - |
| Wild | 1.376 (0.750, 2.526) | 0.360 | 0.866 (0.417, 1.797) | 0.698 |
| **KRAS** | | | | |
| Mutated | 1(ref) | - | 1(ref) | - |
| Wild | 0.611 (0.407, 0.918) | 0.018 * | 0.712 (0.501, 1.013) | 0.059 . |
| **Lympho-vascular Invasion** | | | | |
| No | 1(ref) | - | 1(ref) | - |
| Yes | 2.225 (1.568, 3.157) | <0.0001*** | 1.123 (0.749, 1.684) | 0.574 |

# Table 4

| Table 4 Univariate and multivariate Cox regression models for overall survival (OS) in TCGA patients. | | | | |
| --- | --- | --- | --- | --- |
| **Univariate Cox for TCGA dataset (OS)** | | | **Multivariate Cox for TCGA dataset (OS)** | |
| Variable | HR(95% CI) | P value | HR(95% CI) | P value |
| **TIL Score** | | | | |
| Low | 1(ref) | - | 1(ref) | - |
| High | 0.464 (0.274, 0.787) | 0.0313 * | 0.525 (0.249, 1.106) | 0.090 . |
| Sex | | | | |
| Male | 1(ref) | - | 1(ref) | - |
| Female | 1.011(1.095,1.670) | 0.960 | 0.885 (0.605,1.296) | 0.532 |
| **Age** | 1.024 (1.006, 1.041) | 0.009 ** | 1.025 (1.008, 1.014) | 0.005 ** |
| pT stage | | | | |
| T1-T3 | 1(ref) | - | 1(ref) | - |
| T4 | 2.745 (1.660, 4.540) | <0.0001 *** | 2.581 (1.577, 4.226) | <0.001*** |
| **pN stage** | | | | |
| N0-N1 | 1(ref) | - | 1(ref) | - |
| N2 | 3.467 (2.008, 5.986) | <0.0001*** | 2.105 (1.266, 3.499) | 0.004 ** |
| MSI | | | | |
| MSS | 1(ref) | - | 1(ref) | - |
| MSI-H | 1.156 (0.628, 2.126) | 0.621 | 1.303 (0.687, 2.471) | 0.418 |
| **Stage** | | | | |
| Stage I-II | 1(ref) | - | 1(ref) | - |
| Stage III-IV | 3.438 (2.250, 5.254) | <0.0001 *** | 1.685 (1.015, 2.797) | 0.044 * |
| BRAF | | | | |
| Mutated | 1(ref) | - | 1(ref) | - |
| Wild | 0.882 (0.452, 1.715) | 0.694 | 0.815 (0.398, 1.672) | 0.577 |
| **KRAS** | | | | |
| Mutated | 1(ref) | - | 1(ref) | - |
| Wild | 0.998 (0.644, 1.547) | 0.992 . | 1.014 (0.675, 1.523) | 0.947 |
| Lympho-vascular Invasion | | | | |
| No | 1(ref) | - | 1(ref) | - |
| Yes | 2.150 (1.421, 3.254) | <0.001 *** | 1.005 (0.636, 1.591) | 0.981 |

# Table 5

| Table 5 Univariate and multivariate Cox regression models for overall survival (OS) in MCO patients. | | | | |
|---|---|---|---|---|
| **Univariate Cox for MCO dataset (OS)** | | | **Multivariate Cox for MCO dataset (OS)** | |
| Variable | HR(95% CI) | P value | HR(95% CI) | P value |
| **TIL Score** | | | | |
| Low | 1(ref) | - | 1(ref) | - |
| High | 0.708 (0.568, 0.882) | 0.005 ** | 0.920 (0.722, 1.192) | 0.555 |
| **Sex** | | | | |
| Male | 1(ref) | - | 1(ref) | - |
| Female | 0.753 (0.620, 0.913) | 0.004 ** | 0.698 (0.571, 0.854) | <0.001 *** |
| Age | 1.020 (1.013, 1.030) | <0.0001*** | 1.031 (1.022, 1.040) | <0.0001*** |
| **pT stage** | | | | |
| T1-T3 | 1(ref) | - | 1(ref) | - |
| T4 | 3.102 (2.418, 3.978) | <0.0001*** | 2.131 (1.721, 2.637) | <0.0001*** |
| **pN stage** | | | | |
| N0-N1 | 1(ref) | - | 1(ref) | - |
| N2 | 3.144 (2.354, 4.198) | <0.0001*** | 1.585 (1.231, 2.040) | <0.001 *** |
| **MSI** | | | | |
| MSS | 1(ref) | - | 1(ref) | - |
| MSI | 0.623 (0.476, 0.816) | 0.004 ** | 0.808 (0.559, 1.169) | 0.258 |
| **Stage** | | | | |
| Stage I-II | 1(ref) | - | 1(ref) | - |
| Stage III-IV | 2.863 (2.346, 3.495) | <0.0001 *** | 1.936 (1.516, 2.472) | <0.0001*** |
| **BRAF** | | | | |
| Mutated | 1(ref) | - | 1(ref) | - |
| Wild | 1.176 (0.911, 1.518) | 0.237 | 1.015 (0.754, 1.366) | 0.921 |
| **KRAS** | | | | |
| Mutated | 1(ref) | - | 1(ref) | - |
| Wild | 0.811 (0.662, 0.993) | 0.0358 * | 0.781 (0.637, 0.956) | 0.0166 * |
| **Lympho-vascular Invasion** | | | | |
| No | 1(ref) | - | 1(ref) | - |
| Yes | 2.249 (1.831, 2.761) | <0.0001*** | 1.524 (1.228, 1.891) | 0.0001 *** |

# Figure Legends

**Figure 1**. The framework for the proposed automated deep learning LinkNet for TIL quantification for CRC patients. A LinkNet-based lymphocyte-segmentation model is developed to detect the cell-level lymphocytes on color-normalized image patches for tumors. A tissue classifier is developed to select the tumor regions from whole-slide H&E images (WSIs). The developed LinkNet-based lymphocyte detector is used to quantify the TILs for the selected patches on the WSI by calculating the overall percentage of the TIL area. L represents the region of lymphocytes in the ith patch, P represents the area of one patch, and N represents the total number of tumor patches in a whole slide image.

**Figure 2**. Risk of disease progression (PFS) or death (OS) according to TILs in CRC. Penalized splines modeling of hazard ratio (HR) for (a) PFS in the TCGA CRC cohort; (b) OS in the TCGA CRC cohort; and (c) OS in the MCO cohort. The shaded area around the blue line represents the 95% confidence interval

**Figure 3**. Kaplan–Meier estimates of progression–free survival (PFS) for CRC patients with TIL-High and TIL-Low in the TCGA cohort (a); and Forest plot of effects of TILs on PFS (TIL-High vs. TIL-Low) based on different subgroups according to risk variables (b).

**Figure 4**. Kaplan–Meier estimates of overall survival (OS) for CRC patients with TIL-High and TIL-Low in the TCGA cohort (a); and Forest plot of effects of TILs on PFS (TIL-High vs. TIL-Low) based on different subgroups according to risk variables (b).

**Figure 5**. Kaplan–Meier estimates of overall survival (OS) for CRC patients with TIL-High and TIL-Low in the MCO cohort (a); and Forest plot of effects of TILs on PFS (TIL-High vs. TIL-Low) based on different subgroups according to risk variables (b).

**Figure 1**

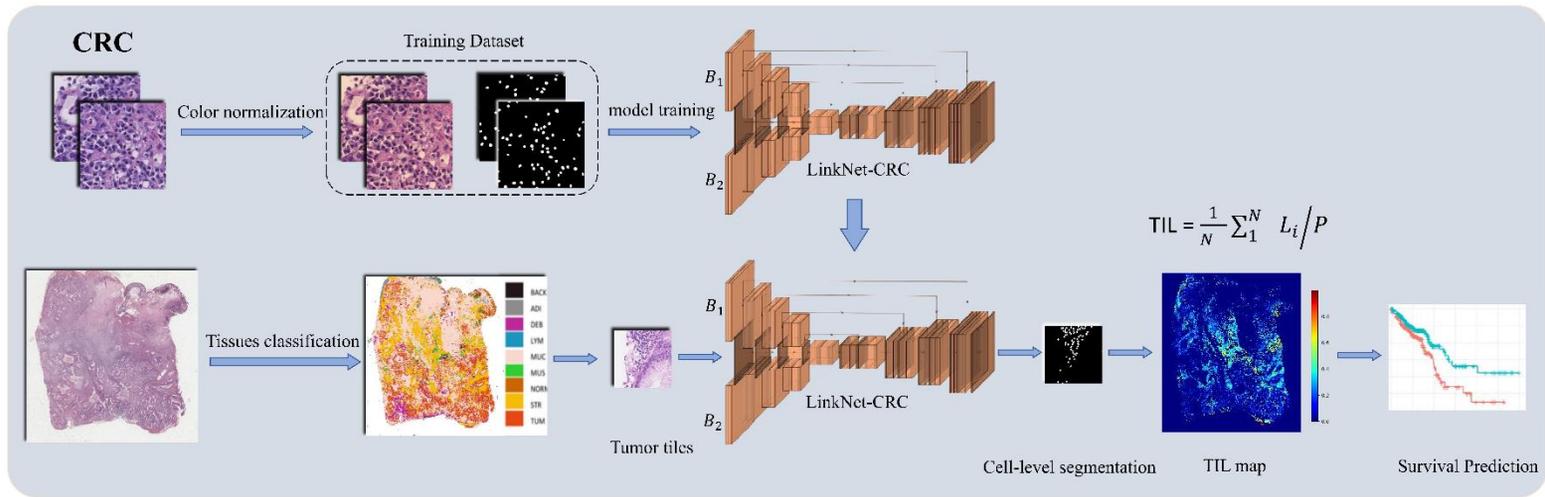

**Figure 2**

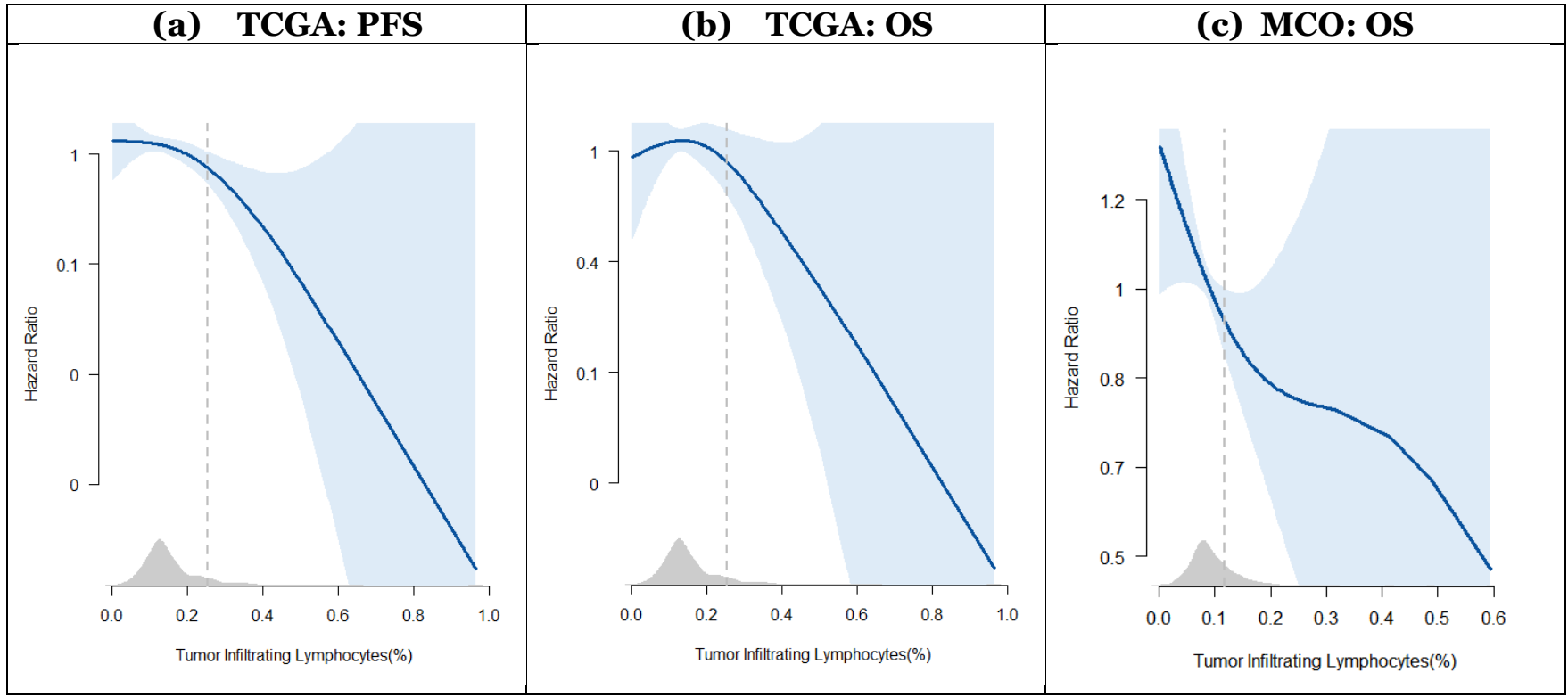

**Figure 3**

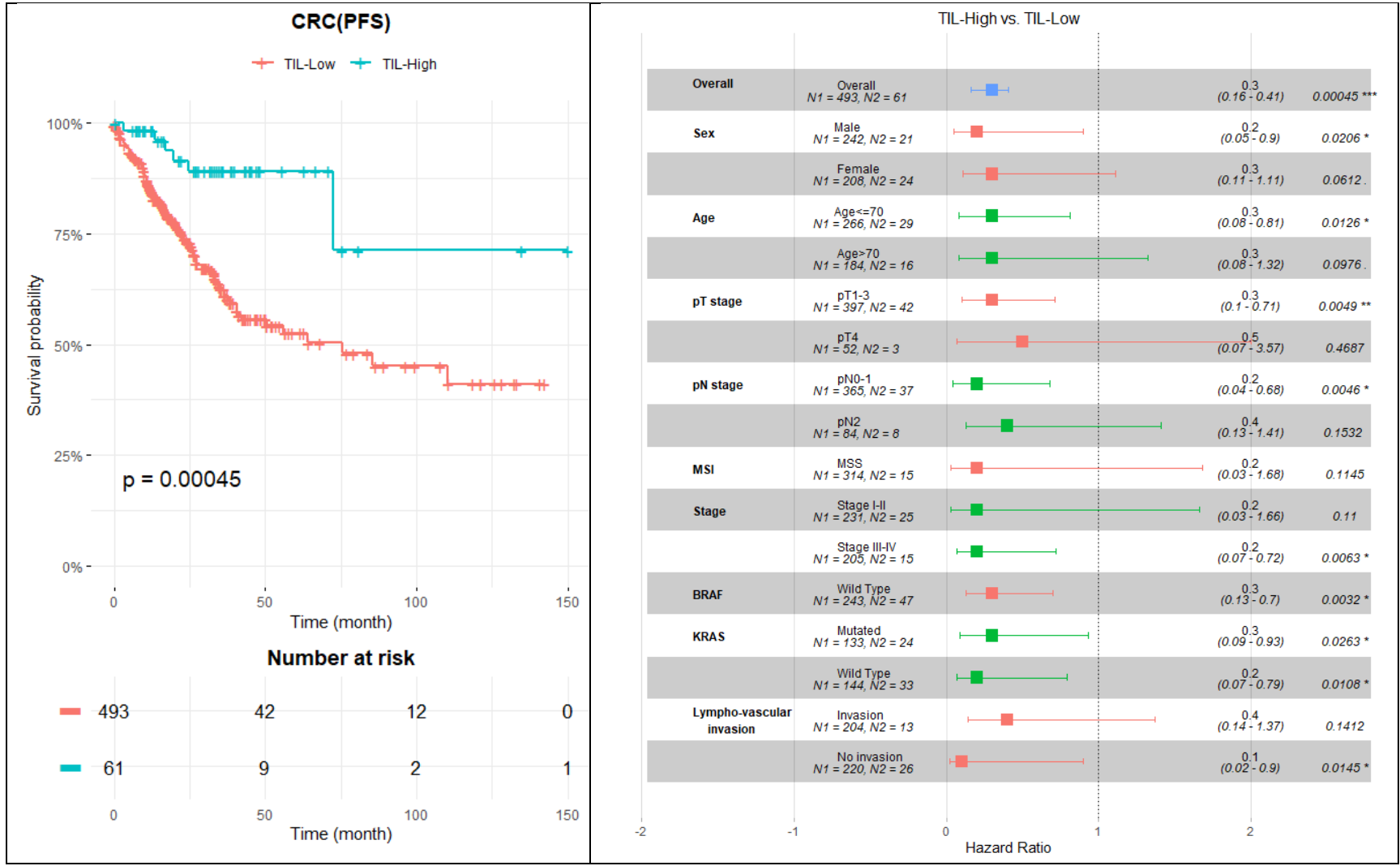

## Figure 4

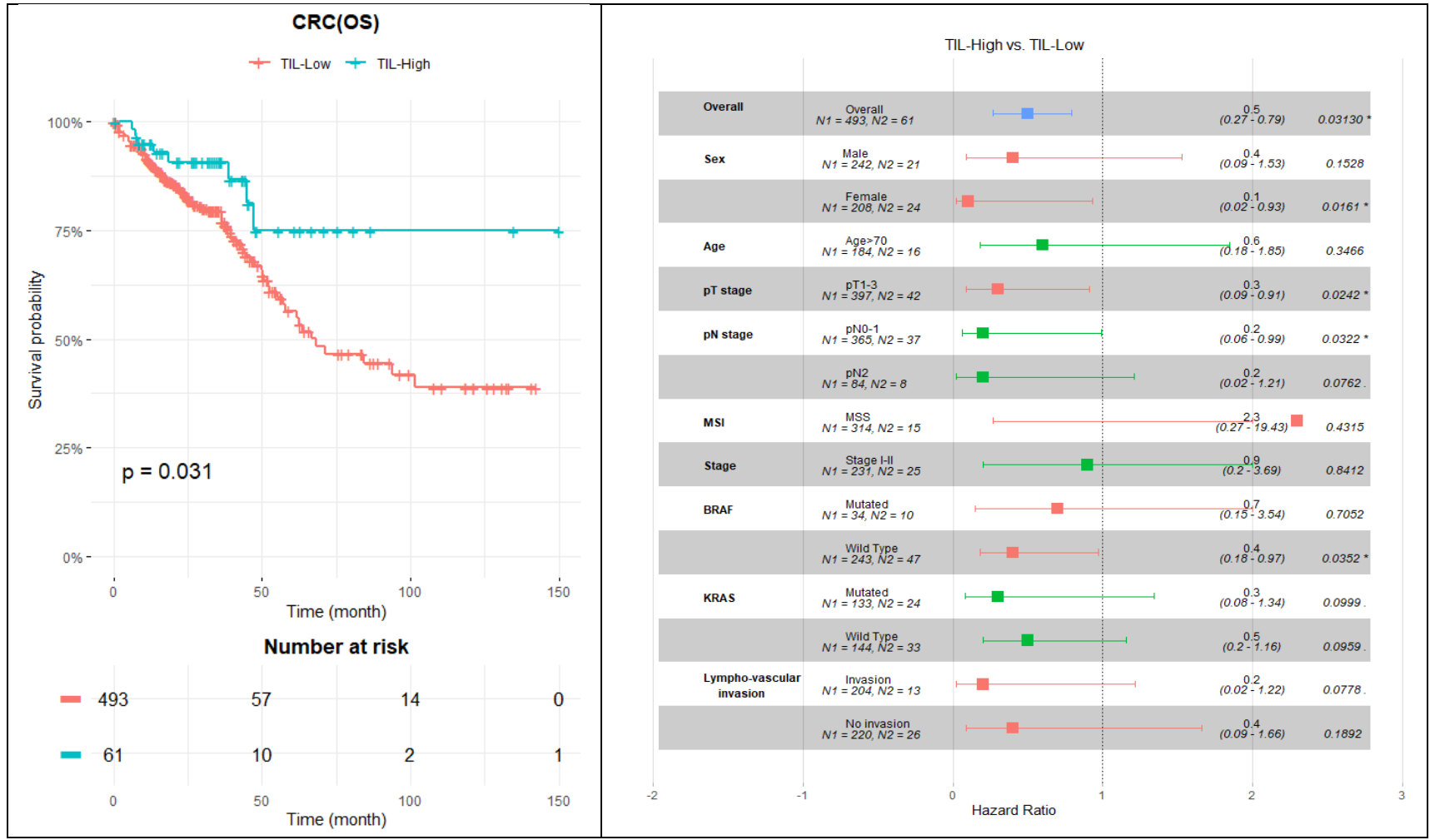

## Figure 5

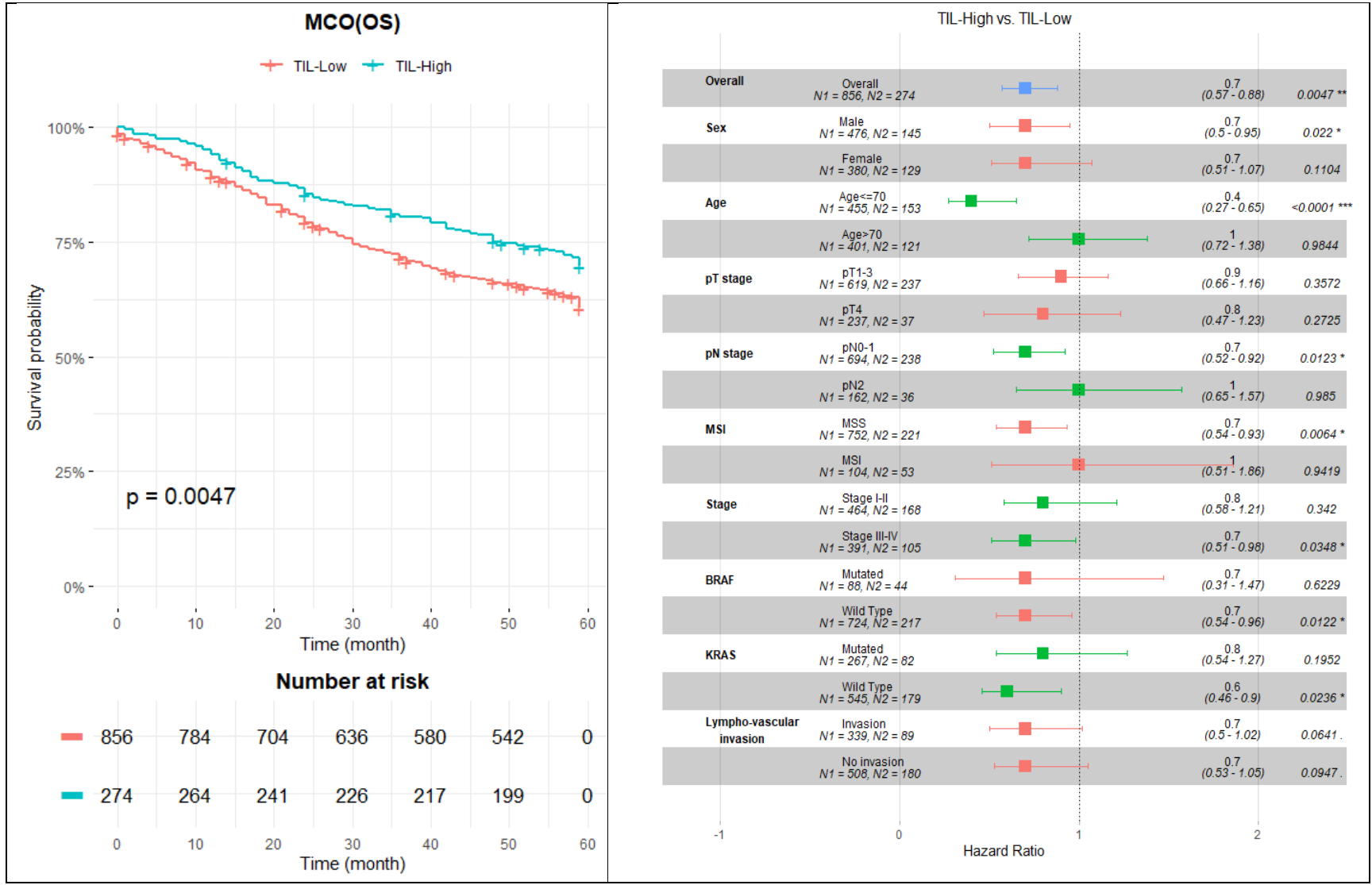